\documentstyle[preprint,eqsecnum,aps,epsf]{revtex}
\newif\iftightenlines\tightenlinesfalse
\tightenlines\tightenlinestrue
\def\eslt{E\llap/_T}
\def\to{\rightarrow}
\def\te{\tilde e}

\def\tb{\tilde b}
\def\tf{\tilde f}
\def\td{\tilde d}
\def\tst{\tilde t}
\def\ttau{\tilde \tau}
\def\tmu{\tilde \mu}
\def\tg{\tilde g}
\def\tnu{\tilde\nu}

\def\tw{\widetilde W}
\def\tz{\widetilde Z}

\begin{document}
%
\preprint{\vbox{\baselineskip=14pt%
   \rightline{FSU-HEP-970501}\break 
   \rightline{UCD 97-12}\break 
   \rightline{APCTP 97-09}
   \rightline{UH-511-870-97}
}}
%
\title{COLLIDER PHENOMENOLOGY FOR SUPERSYMMETRY WITH 
LARGE $\mathop{\bf tan}\beta$}
\author{Howard Baer$^1$, Chih-hao Chen$^2$, Manuel Drees$^3$, 
Frank Paige$^4$\\
 and Xerxes Tata$^{5}$}
\address{
$^1$Department of Physics,
Florida State University,
Tallahassee, FL 32306 USA
}
\address{
$^2$Davis Institute for High Energy Physics,
University of California,
Davis, CA 95616 USA
}
\address{
$^3$APCTP, 207-43 Cheongryangri-dong,
Tongdaemun-gu, Seoul 130-012
Korea
}
\address{
$^4$Brookhaven National Laboratory, 
Upton, NY 11973 USA
}
\address{
$^5$Department of Physics and Astronomy,
University of Hawaii,
Honolulu, HI 96822 USA
}
\date{\today}
\maketitle
\begin{abstract}

If the parameter $\tan\beta$ of the minimal supersymmetric model is
large, then $b$ and $\tau$ Yukawa interactions are important. These
can significantly modify the masses and decays of sparticles.
We describe new
calculations which allow a reliable exploration of large $\tan\beta$ values,
and discuss implications for collider experiments. 
For large values of $\tan\beta$, 
charginos and neutralinos may dominantly decay to
$\tau$-leptons or $b$-quarks. The usual cross sections for multilepton
signatures may be greatly reduced, but SUSY may be detectable via new
signals involving $\tau$'s or $b$'s in the final state.

\end{abstract}

\medskip

\pacs{PACS numbers: 14.80.Ly, 13.85.Qk, 11.30.Pb}


Weak scale supersymmetry \cite{REV} 
is one of the most promising candidates for physics beyond the
Standard Model (SM). 
The minimal supersymmetric model (MSSM) provides 
a well-motivated framework for investigations of the
experimental consequences of weak scale supersymmetry.
The MSSM is essentially the supersymmetrized version of
the Standard Model (SM), with $R$-parity
assumed to be conserved.
Supersymmetry breaking 
is incorporated by including soft supersymmetry-breaking
interactions consistent with the assumed symmetries. 
Within the minimal supergravity (mSUGRA) framework, which has been
adopted for many phenomenological analyses, the large number of
independent soft breaking
parameters are related by renormalization group (RG) evolution
to universal scalar and gaugino
masses ($m_0$ and $m_{1/2}$) and one universal trilinear coupling ($A_0$), all
specified at the scale $M_X \sim 2 \times 10^{16}$~GeV; these, together
with $\tan\beta = {v_u}/{v_d}$ and the sign of the Higgsino mass 
parameter $\mu$ completely specify the
model. 

In many studies of SUSY phenomenology
involving cascade decays of sparticles at colliders,
only small to moderate values of $\tan\beta$, i.e.,
$\tan\beta\sim1\hbox{--}10$, have been 
considered\cite{REV}. Partly, this was because event generation programs such 
as ISAJET\cite{ISAJET} did not include all
effects of bottom and tau Yukawa couplings.
The third generation Yukawa couplings are given by
\begin{eqnarray*}
f_t={g m_t\over\sqrt{2}M_W\sin\beta},\ f_b={g m_b\over\sqrt{2}M_W\cos\beta},
\ f_{\tau}={g m_{\tau}\over\sqrt{2}M_W\cos\beta}.
\end{eqnarray*}
As $\tan\beta$ becomes large, $f_b$ and $f_\tau$ become large and 
comparable in strength to $f_t$.
In the MSSM, the parameter $\tan\beta$ can
range as high as $\tan\beta\simeq 45\hbox{--}50$ 
before running into various theoretical and/or experimental constraints. 
Such large $\tan\beta$ values are
actually preferred in some $SO(10)$ GUT models with Yukawa 
unification. 

Large values of $\tan\beta$ affect SUSY phenomenology and impact upon 
the search for 
weak scale supersymmetry at collider experiments in several ways:
\begin{itemize}

\item Large $b$ and $\tau$ Yukawa couplings contribute negatively to the
RG running of the $\tb_{L,R}$ and $\ttau_{L,R}$ soft
masses.
In models with a common scalar mass at some high scale, third generation
scalar masses will be driven
to lower values than masses for the corresponding 
first and second generation squarks and sleptons.
In addition, left-right mixing of stau and sbottom 
eigenstates can cause a further reduction in the $\ttau_1$ and $\tb_1$
masses.

\item It is well known that the large top Yukawa coupling drives the Higgs
squared mass $m_{H_u}^2$ to negative values, resulting in radiative 
breakdown of electroweak
symmetry. At large $\tan\beta$, the large $b$ and $\tau$ Yukawa couplings
drive the other Higgs squared mass $m_{H_d}^2$ to small or negative values
as well. This results \cite{DN}
overall in a {\it decrease} in mass for the pseudo-scalar
Higgs $m_A$ relative to its value at small $\tan\beta$. Since the values of the
heavy scalar and charged Higgs boson masses are related to $m_A$, 
these are also reduced.

\item  For large values of $\tan\beta$, $b$ and $\tau$ Yukawa couplings become
comparable in strength to the usual gauge interactions, so that Yukawa
interaction 
contributions to sparticle decay rates are non-negligible and can even
dominate. This could manifest itself as lepton non-universality in SUSY
events. Also, because of the reduction in masses referred to
above, chargino and neutralino decays to stau, sbottom
and various Higgs bosons 
may be allowed, even if the corresponding decays would be 
kinematically forbidden for small $\tan\beta$ values. 
The reduced stau/sbottom/Higgs masses can also
increase sparticle branching ratios to third generation particles 
via virtual effects. These enhanced decays to third generation
particles can radically alter \cite{THIRD} 
the expected SUSY signatures at colliders.


\item Tau Yukawa interactions can alter the mean polarization of the
$\tau$'s produced in chargino and neutralino decays. This, in turn, alters
the energy distribution of the visible decay products of the $\tau$,
and hence the efficiency with which the $\tau$ signals can be detected.

\end{itemize}

These considerations motivated us to begin a systematic exploration
of how signals for supersymmetry may be altered if $\tan\beta$ indeed
turns out to be very large. To facilitate this analysis, we have made
several new calculations (described below) and incorporated these into
ISAJET 7.28. ISAJET allows\cite{ISASUG}
the simulation of supersymmetry for either a
general set of input (weak scale) parameters or for the more restrictive mSUGRA
parameter set from which these weak scale parameters are computed.
Here, we will only discuss the
improvements that we have made in ISAJET to allow a reliable analysis
for large values of $\tan\beta$.

A first step is to obtain reliable predictions for superpartner and Higgs
boson masses as a function of parameters.  For both MSSM and mSUGRA 
parameter sets, we have included
bottom squark and tau slepton mixing effects. 
We found that the pseudoscalar mass $m_A$, obtained using the 1-loop
effective potential, is unstable by up to factors of two 
against scale variations for relatively low values of scale choice
$Q\sim M_Z$. 
This instability would be presumably corrected by inclusion of 
2-loop corrections. 
We find the choice of scale
$Q\sim\sqrt{m_{\tst_L}m_{\tst_R}}$ to empirically yield stable predictions of 
Higgs boson masses in the RG improved 1-loop effective potential
(where we include contributions from all third generation 
particles and sparticles).
This scale choice effectively includes some
important two loop effects, and yields predictions for light scalar Higgs boson
masses $m_h$ in close accord with the results of Ref. \cite{CARENA}.

We illustrate the $\tan\beta$ dependence of superpartner and Higgs boson masses
in Fig.~1, where we have chosen mSUGRA parameters 
($m_0,\ m_{1/2},\ A_0$)=(150,150,0) GeV. There is little
variation in the mass of the $\td_L$ squark, the $\te_R$ slepton or the 
neutralino $\tz_1$. The $\tst_1$ and $\tw_1$ mass varies mainly at 
low $\tan\beta$. However, there is a significant decrease in the $\tb_1$
and $\ttau_1$ masses as $\tan\beta$ increases. The mass decrease is so severe
that for $\tan\beta\sim 38$, $m_{\tw_1}>m_{\ttau_1}$ so that the
two body decays $\tw_1\to\ttau_1\nu_{\tau}$ and $\tz_2\to \ttau_1\tau$
become kinematically accessible. Being the only allowed two body
modes, these dominate chargino and neutralino decays.
Even more noteworthy is the drastic decrease in the pseudoscalar Higgs mass
$m_A$, which drops from $m_A\sim 400$ GeV at $\tan\beta =2$ to less than
40 GeV for $\tan\beta\sim 47$. For even larger values of $\tan\beta$,
$m_A^2<0$, so that electroweak symmetry breaking is not correctly obtained. 
A slightly stronger upper bound on $\tan\beta$
comes from non-observation of $Z\to hA$ events at the $10^{-4}$ level in
experiments at LEP\cite{HALEP}.
For these very high values of $\tan\beta$, the $H$ and $H^\pm$ masses also
decrease, so that $t\to bH^+$ can occur, potentially in conflict with
limits\cite{CDFHPLUS}
from the CDF experiment. Finally, we note that there is a (hard to see) kink in
the $h$ mass curve at $\tan\beta =43$. Here, the $H$ and $h$ masses become
equal, and for larger $\tan\beta$, these particles switch roles, so that the 
light Higgs $h$ mass decreases along with the curve for the
pseudoscalar $A$. 

The next step is to properly evaluate the various 
sparticle and Higgs boson branching fractions for large $\tan\beta$.
We have, therefore,
re-calculated the branching fractions for the $\tg$, $\tb_i$, $\tst_i$,
$\ttau_i$, $\tnu_{\tau}$, $\tw_i$, $\tz_i$, $h$, $H$, $A$ and $H^\pm$
particles and sparticles including sbottom and stau mixing as well as
effects of $b$ and $\tau$ Yukawa interactions. 
For Higgs boson decays, we use the formulae in Ref. \cite{BISSET}.
We have recalculated the decay widths for 
$\tg\to tb\tw_i$ and $\tg\to b\bar{b}\tz_i$. These have been calculated previously
by Bartl {\it et.\ al.}\cite{bartl}; our results agree with theirs if we
use pole fermion masses to calculate 
the Yukawa couplings. In ISAJET, we use the
running Yukawa couplings evaluated at the scale $Q=m_{\tg}$ ($m_t$) to compute
decay rates for the gluino ($\tw_i$,$\tz_i$). This seems a more
appropriate choice, and it significantly alters
the decay widths when effects of $f_b$ are important.
The $\tz_i\to \tau\bar{\tau}\tz_j$ and $\tz_i\to b\bar{b}\tz_j$
decays take place via eight diagrams ($\tf_{1,2}$,
$\bar{\tf}_{1,2}$, $Z$, $h$, $H$ and $A$ exchanges). We have 
calculated these decays (neglecting $b$ and $\tau$ masses except
in the Yukawa couplings and in the phase space integration). 
We have also computed
the widths for decays $\tw_i\to\tz_j \tau\nu$ which are mediated by
$W$, $\ttau_{1,2}$, $\tnu_{\tau}$ and $H^{\pm}$ exchanges. 

To illustrate the importance of the Yukawa coupling effects, 
we show selected branching ratios of the
$\tw_1$, $\tz_2$ and $\tg$ in Fig.~2{\it a--c} respectively. 
In all frames we take $\mu >0$.
Frames {\it a})
and {\it b}) are for the mSUGRA case
($m_0,\ m_{1/2},\ A_0 )=(150,150,0)$ GeV.
For low $\tan\beta$
we see that the $\tw_1\to e\nu\tz_1$ and $\tw_1\to\tau\nu\tz_1$ 
branching ratios are very close in magnitude, reflecting the smallness
of $f_{\tau}$. For $\tan\beta \agt 10$, these 
branchings begin to diverge, with the branching to $\tau$'s 
becoming increasingly 
dominant. For $\tan\beta >40$, the two body mode $\tw_1\to \ttau_1\nu$
opens up and quickly dominates. Since this decay
is followed by $\ttau_1\to \tau\tz_1$, the end product of chargino 
decays here are almost exclusively 
tau leptons plus missing energy.

In frame {\it b}), we see at low $\tan\beta$ the $\tz_2\to e\bar{e}\tz_1$ and
$\tz_2\to \tau\bar{\tau}\tz_1$ branchings are large ($\sim 10\%$) and equal,
again because of the smallness of the Yukawa coupling.
Except for parameter regions where the leptonic decays of $\tz_2$ are
strongly suppressed, $\tw_1\tz_2$ production
leads to the well known $3\ell$ ($=e,\mu$) signature for the 
Tevatron and LHC colliders\cite{REV}. 
As $\tan\beta$ increases beyond 10, these 
branchings again diverge, and increasingly $\tz_2\to\tau\bar{\tau}\tz_1$
dominates. Also, we see that the $\tz_2\to b\bar{b}\tz_1$ branching fraction
becomes increasingly dominant for large $\tan\beta$.
For $\tan\beta >40$, 
$\tz_2\to \tau\ttau_1$ opens up, and becomes quickly close to 100\%. Near the
edge of parameter space ($\tan\beta \sim 45$), the $\tz_2\to \tz_1 h$ decay
opens up, resulting in a reduction of the $\tz_2\to \tau\ttau_1$
branching fraction. 

Frame {\it c}) shows several gluino branching fractions for 
$(m_0,\ m_{1/2},\ A_0 )=(700,250,0)$~GeV, for which 
$m_{\tg}\simeq 700$~GeV. For $\tan\beta\sim 2$, the 
$\tg\to t\bar{t}\tz_1$ and $\tg\to tb\tw_1$ branching fractions
dominate the decay. As $\tan\beta$ increases, $|\mu |$ decreases so that
$\tg$ decays into heavier chargino and neutralino states become allowed, 
and more cascading takes place in the $\tg$ decays. We also see from 
frame {\it c}) that as $\tan\beta$ increases, 
ultimately the $\tg\to b\bar{b}\tz_i$ branching fraction becomes dominant
until for very large $\tan\beta$ the two body mode $\tg\to \tb_1 b$ opens up.
For $\tan\beta\sim 40$, the $\tg\to b\bar{b}\tz_i$ branching fraction occurs
at 4, 14, 11 and 6\% for $\tz_i=\tz_1,\ \tz_2,\ \tz_3$ and $\tz_4$, 
respectively. 
We also see from Fig.~2 that for $\tan\beta \alt 10$, the $\tau$ and $b$
Yukawa coupling effects are small, and conclusions from previous SUSY analyses
are still valid.   

At $e^+e^-$ colliders, it is possible the $\ttau_1\bar{\ttau_1}$ production
could be 
accessible to experiments, whereas $\tmu\bar{\tmu}$ and $\te\bar{\te}$ would be
inaccessible. In this case, SUSY would be revealed as
acollinear $\tau$-pair events. The energy spectrum of the hadronic decay
products of the $\tau$ lepton
depends on the helicity of the $\tau$, so that its measurement could
yield information 
on the left-right mixing of the parent, and hence on $A_{\tau}$ and
$\tan\beta$ \cite{NOJIRI}. For this reason (and also
because the detection efficiency for taus at hadron colliders is
sensitive to it), we include in ISAJET
a calculation of tau polarization resulting from various possible parents:
$\tz_i$, $\tw_i$, $\ttau_i$, $\tnu_\tau$ and $H^\pm$. Exact matrix 
elements are used for tau lepton decays in ISAJET. Since ISAJET does not 
include chargino or neutralino spin correlations, an average polarization is 
computed for three body decays from these sources. This polarization is
used in the generation of subsequent decays of the $\tau$ lepton.

If $\tw_1\overline{\tw_1}$ pair production is accessible, 
then for large $\tan\beta$
a non-universality in $e/\mu /\tau$ content of signal events might be
observed. The angular distribution of taus should make it possible to
distinguish between chargino and stau events, even if other signals
such as $\ell+\tau+\eslt$ or $j+\tau+\eslt$ are suppressed.
Likewise, if $\tz_1\tz_2$ is the only reaction accessible, enhanced decays
to $b$'s or $\tau$'s may be evident. Finally, for mSUGRA models with low
values of $\tan\beta$, the $H,\ A$ and $H^{\pm}$ are expected to be very
heavy. 
An intriguing prospect at large $\tan\beta$ is that, 
since the additional Higgs bosons can be quite light, that $hZ$, $HZ$,
$hA$, $HA$ and
$H^\pm H^\mp$ might all be at least kinematically accessible. 
The neutral Higgs bosons will 
usually decay to $b\bar{b}$ pairs, while for the charged Higgs boson,
the decay $H^{\pm} \to \tau\nu_{\tau}$ frequently dominates.

Turning to hadron colliders, we recall that the best bound on gluino and
squark masses comes from an analysis of ${\rm jets}+\eslt$ events \cite{ETMISS}
at the
Fermilab Tevatron; 
very similar bounds have been obtained from analyses of 
jets$+2\ell+\eslt$ events\cite{DIL}.
For $\tan\beta \alt 10$, the isolated trilepton
signal from $\tw_1\tz_2\to 3\ell$ potentially offers the greatest SUSY reach
at the integrated 
luminosity\cite{REV} that should be available at the Main
Injector. 
As $\tan\beta$ grows, we see from Fig.~2 that 
the branching fraction of $\tw_1$ and especially $\tz_2$ into $e$'s and
$\mu$'s diminishes greatly. Thus, the clean $3\ell$ signal is considerably
reduced from its value for low values of $\tan\beta$ examined in
earlier studies. However, a host of new signal channels can potentially
be used for large $\tan\beta$ studies. Event topologies such as $\ell\tau\tau$,
$\ell\bar{\ell}\tau$, $bb\tau$, $b\tau\tau$ and $3\tau$ (with, possibly,
like-sign and like-flavour $\ell$ or $\tau$ pairs in the event)
are possibilities 
for Fermilab Tevatron experiments accumulating $\agt 2$ fb$^{-1}$ of 
integrated luminosity. 

We have calculated the Tevatron signal and background in the $\ell\tau\tau$ and
$\ell\bar{\ell}\tau$ 
channels for the mSUGRA point chosen in 
Fig.~2{\it a} and {\it b} with $\tan\beta =45$.
Already with just basic acceptance cuts,
we found the combined 
signal (17~fb) to be obervable at $5\sigma$ level above SM background (28~fb) 
provided $\agt 2.4$~fb$^{-1}$
of integrated luminosity is collected. In our analysis, a $\tau$ is 
identified if it is a 1 or 3 isolated charge prong 
hadronic jet with $E_T>15$ GeV, $M<1.8$~GeV and net charge of $\pm 1$. 
Isolated leptons are required to
have $p_T>10$~GeV; some $\ell$, $j$ and $\eslt$ trigger requirements are also
imposed. SM backgrounds from  $t\bar{t}$, $W + j$, $Z + j$, $WW$ and
$WZ$ were included in our assessment of the signal, and QCD jets with
$E_T=15$ ( $\geq 50$)~GeV were
misidentified as $\tau$'s 0.5\% (0.1\%) of the time\cite{CDFTAU}.
These calculations indicate that the $\ell\tau\tau$ and $\ell\bar{\ell}\tau$
signals should be observable for some regions of parameter space where
the canonical $3\ell$ signal fails.
Hence, we strongly urge our experimental colleagues to keep in mind that SUSY
signals at large $\tan\beta$ might be considerably different 
from general expectations:  
{\it identification of signals with isolated $\tau$'s should be a 
priority for Run 2}\cite{FN1}.

At the CERN LHC, observable SUSY signals at low $\tan\beta$ values
are expected in various multijet plus multilepton plus $\eslt$ channels.
For large 
$\tan\beta$, rates for many of these leptonic signals are expected to decrease,
but as for the Tevatron, rates for events containing $b$-jets and
isolated hadronic $\tau$'s should increase, partly due to
enhanced production of Higgs bosons $H$ and $A$ in gluino
and squark cascade decays. 

In summary, if $\tan\beta$ is large, SUSY may manifest itself via events
rich in $\tau$ leptons and $b$-jets; signals in conventional leptonic channels
may be much reduced! The detection of these novel signals, especially at
hadron colliders, could require
the development of new algorithms for efficiently identifying isolated $\tau$
leptons, possibly in events containing several jets.

%
\acknowledgments
We thank A. Bartl, M. Bisset, D. Casta\~no and W. Porod for 
valuable discussions.
This research was supported in part by the U.~S. Department of Energy
under contract numbers DE-FG05-87ER40319, DE-FG03-91ER40674, 
DE-AC02-76CH00016, and DE-FG-03-94ER40833. 
%
%

\newpage
%
%

\iftightenlines\else\newpage\fi
\iftightenlines\global\firstfigfalse\fi



\begin{figure}
\iftightenlines\epsfxsize=5in
\centerline{\epsfbox{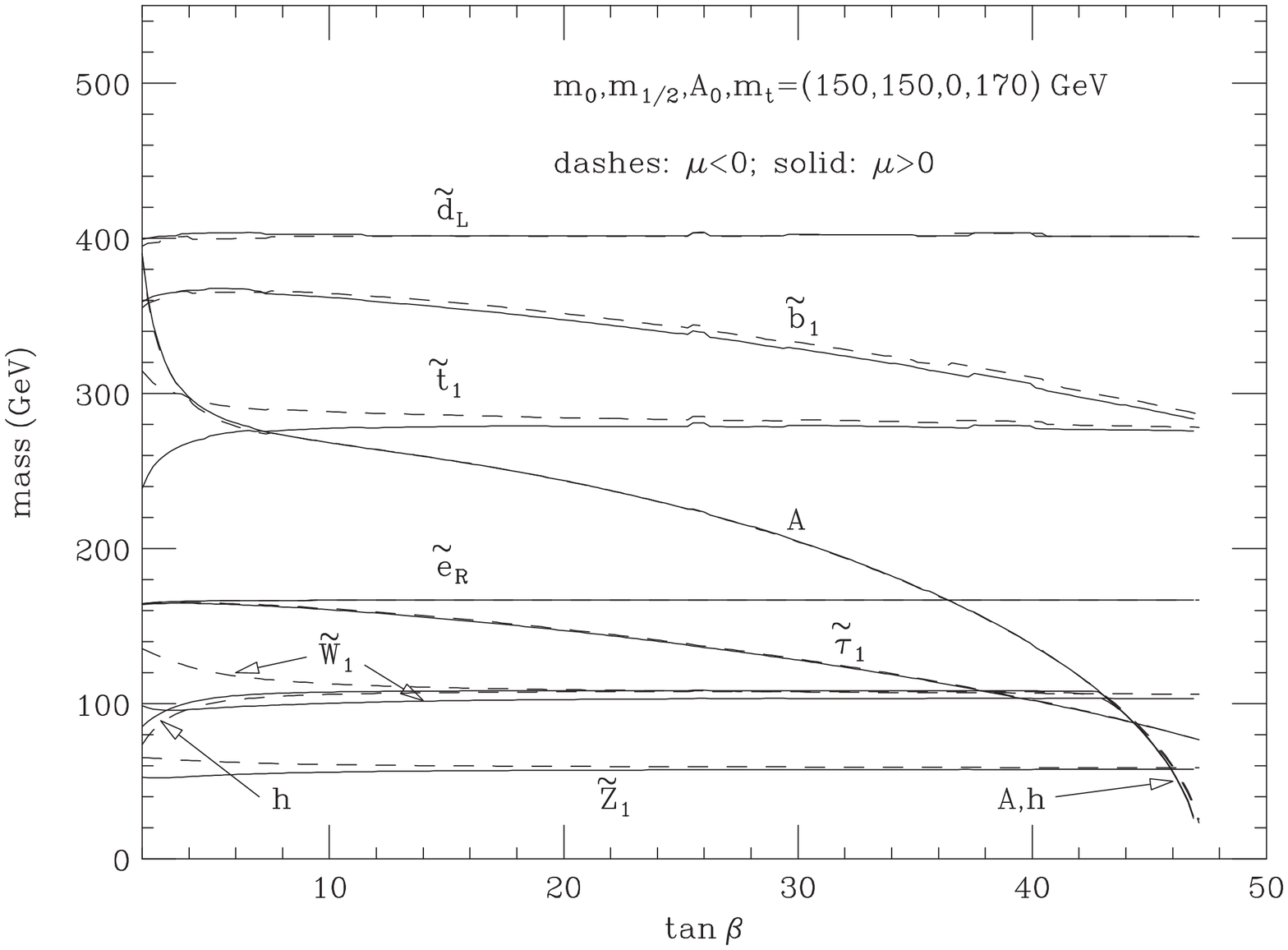}}
\medskip\fi
\caption[]{A plot of various sparticle and Higgs masses 
versus $\tan\beta$ for the 
parameters ($m_0,m_{1/2},A_0)=(150,150,0)$ GeV, for both signs of the 
parameter $\mu$. We take $m_t=170$ GeV.}
\label{FIG1}
\end{figure}

\begin{figure}
\iftightenlines\epsfxsize=6.25in
\centerline{\epsfbox{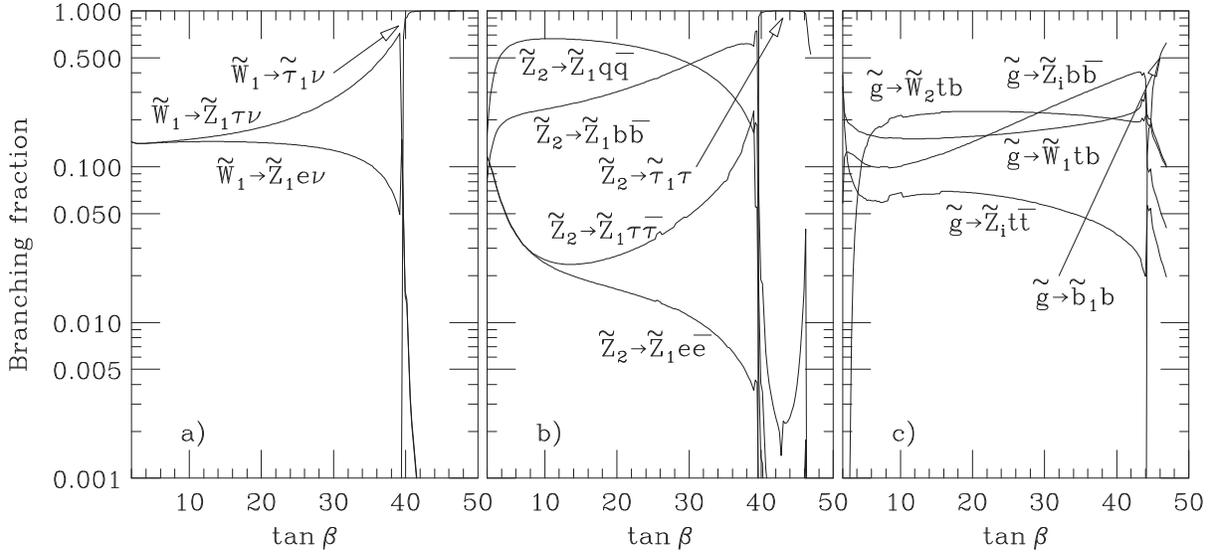}}
\medskip\fi
\caption[]{A plot of sparticle branching fractions versus $\tan\beta$. In 
{\it a}) and {\it b}), we take the parameters 
($m_0,m_{1/2},A_0)=(150,150,0)$ GeV while {\it c}) uses 
($m_0,m_{1/2},A_0)=(700,250,0)$ GeV. In all frames, $\mu >0$ and 
$m_t=170$ GeV.}
\label{FIG2}
\end{figure}


\begin{references}
\bibitem{FN1} For this parameter point, we have also checked that with
the cuts of H. Baer {\it et. al.}, 
Phys. Rev. D{\bf 54}, 5866 (1996) and using $E_T^c=100$~GeV, the
multijet plus $\eslt$ signal is insensitive to $\tan\beta$.
\end{references}
\end{document}